\documentclass[prl]{revtex4}%
\pdfoutput=1

\usepackage{epsfig}
\usepackage{graphics}

\begin{document}

\title{Indeterminacy of Holographic Quantum Geometry}

\author{Craig J. Hogan}
\affiliation{University of Chicago and Fermilab}

\begin{abstract} 
An effective theory   based on wave optics  is used to describe indeterminacy of position in holographic spacetime with a UV cutoff at the Planck scale.  Wavefunctions  describing spacetime positions     are modeled as complex disturbances  of quasi-monochromatic radiation.   It is shown that the product of standard deviations of two position wavefunctions in the plane of a holographic light sheet  is  equal to the product of their normal separation and the Planck length.  For macroscopically separated positions the transverse uncertainty is much larger than the Planck length, and is predicted to be observable  as a  ``holographic noise" in relative position with a distinctive shear spatial character, and an absolutely normalized frequency spectrum with no parameters once the fundamental wavelength is fixed from the theory of gravitational thermodynamics.  The spectrum of holographic noise is  estimated  for the  GEO600 interferometric gravitational-wave detector, and is shown to approximately account for currently unexplained noise  between about 300 and 1400Hz. In a holographic world, this result directly and precisely measures the fundamental minimum interval of time. 
\end{abstract}
\maketitle
\section{introduction}
Theory suggests that spacetime is holographic--- that quantum geometry has a native formulation on light sheets, or two-dimensional geometrical wavefronts, with a UV cutoff or minimum length at the Planck scale\cite{'tHooft:1993gx,Susskind:1994vu,Jacobson:1995ab,Banks:1996vh,'tHooft:1999bw,Bousso:2002ju,Padmanabhan:2006fn,Padmanabhan:2007en}. This paper proposes a  quantitative formalism to calculate the properties of a  new observable phenomenon of  quantum  holographic geometry,  ``holographic noise''.  An effective theory of quantum geometry adapts the theory of correlations in  wave optics\cite{bornwolf}  to calculate correlations of   position wavefunctions,   the resulting indeterminacy  in transverse position, and the spectrum of the noise.   The method used here
improves earlier derivations\cite{Hogan:2007rz,Hogan:2007hc,Hogan:2007ci,Hogan:2007sw,Hogan:2007pk,Hogan:2008zw}:  it manifestly respects Lorentz symmetry, and provides a controlled calculation  of all steps connecting  the fundamental holographic minimum time interval   with the observable noise spectrum.

The spatial correlation of the noise has a distinctive transverse character connected with
its holographic origin, and the absolute normalization of the noise is fixed by gravitational thermodynamics.
The precisely predicted spectrum and specific transverse  spatial character   provide zero-parameter experimental signatures that distinguish holographic noise from a variety of other hypothetical types of metric indeterminacy and noise\cite{ellis,AmelinoCamelia:1998ax,AmelinoCamelia:1999gg,AmelinoCamelia:2001dy,AmelinoCamelia:2003zf,AmelinoCamelia:2005qa,Smolin:2006pa}.  These holographic features appear to be  not specific to the particular wave model used here to compute the effect, but to be general properties of uncertainty and nonlocality associated with  holographic geometry with an ultraviolet cutoff at the Planck scale.

A simple model is used to describe the effect of holographic noise in a Michelson interferometer.  The response of an interferometric gravitational-wave detector\cite{hough} is estimated in a form that allows for direct comparison with the quoted levels of instrument noise in terms of equivalent gravitational wave metric strain.  The holographic noise spectrum estimated here agrees  with  currently unexplained noise in the GEO600 detector\cite{GEO} in its most sensitive range of frequencies, between about 300 and 1400Hz.  If holographic noise is indeed the explanation, this result represents a direct measurement of the fundamental maximum frequency, or minimum interval of time.

\section{wave theory of transverse position}

The standard  framework and nomenclature  of wave optics\cite{bornwolf} maps directly onto wave mechanics of a holographic geometry.  The system is represented as a field of complex oscillation amplitudes of quasi-monochromatic ``radiation'' of wavelength $\lambda$ that propagates in flat spacetime via the wave equation.   This carrier radiation is not  an ordinary physical field, but its complex amplitudes are used to describe correlations of  wavefunctions encoding spatial positions  in  a holographic model of spacetime.  

Geometrical wave surfaces or geometrical wavefronts   correspond to families of two dimensional light sheets, locally orthogonal to a nominal  propagation direction. The direction is determined in the quantum geometry interpretation by an experimental measurement setup, such as an interferometer beam or optical surface. Position wavefunctions along this direction have a width given by $\approx \lambda$ and do not enter into the following discussion.
A ``complex disturbance''   $\psi(x,y,t)$ defined on a surface $S$ corresponds to a  wavefunction for a transverse event position.
The wavefunction 
$\psi'$ on another surface $S'$  is linearly propagated from  that  on $S$ via the wave equation.
(In the following, primed quantities all refer to events on  $S'$.)
The  ``intensity''   $I=\langle \psi^*\psi\rangle$ at one point  corresponds to a time-averaged squared modulus of the quantum-mechanical amplitude to be at that position, identified with  the  probability density for event position.
  The optical terms ``mutual intensity''  $J$ and ``complex degree of coherence''  $\mu$ correspond to   correlation operators in  quantum mechanics.    Measurement of the position of a body corresponds to an interaction that reduces the wave amplitude $\psi$ that represents the spacetime to an eigenstate of transverse position.  Because wave optics describes the propagation of light, the spacetime transformation properties of the correlations are automatically consistent with Lorentz invariance.

Consider two points $P_1'$ and $P_2'$ on surface $S'$, corresponding to pointlike events.
The disturbance at point $P_1'$  on surface $S'$ due to  a disturbance   from a small surface element $m$ on  surface $S$ is
\begin{equation}
\psi'_{m1}(t)=A_m(t-L_{m1}) e^{-2\pi i  (t-L_{m1})/\lambda}/L_{m1}
\end{equation}
where $L_{m1}$ is the separation of the point $P_1'$ from $m$, $\lambda$ is the wavelength, and the speed of light $c$ is taken to be unity. Here, $|A_m|$ denotes the amplitude and arg $A_m$ the phase of ``radiation'' from the $m$th element.  An exactly similar expression applies for $P_2'$.
The contribution  from the one patch $m$  to the time-averaged wavefunction correlation at the two points on $S'$ is given by 
 \begin{equation}
\langle \psi'_{m1}(t) \psi^{'*}_{m2}(t) \rangle
=\langle A_{m}(t-L_{m1}) A_{m}^{*}(t-L_{m2}) \rangle
e^{2\pi i (L_{m1}-L_{m2})/\lambda}/L_{m1}L_{m2}
\end{equation}
Assuming $L_{m1}-L_{m2}$ is small compared with the  coherence length of the carrier, we neglect the distance difference in the time averaged argument of $A$ to obtain
\begin{equation}
\langle \psi'_{m1}(t) \psi_{m2}^{'*}(t) \rangle=\langle A_{m}(t) A_{m}^{*}(t) \rangle e^{2\pi i (L_{m1}-L_{m2})/\lambda}/L_{m1}L_{m2}.
\end{equation}
The overall correlator of the wavefunction (in optical nomenclature, the ``mutual intensity'' $J_{12}'$) is given by summing over the contributions to $\psi'$ at $P_1'$ and $P_2'$ from all patches $m,n$ on surface $S$.
Assume an incoherent ``source'' on surface $S$ and thereby neglect correlations in the source, so contributions in the sum with $m\neq n$ average to zero. (Quantum mechanically, this corresponds to a classically defined prepared state without fine scale phase correlations.) Then combining these expressions we write the sum as an  integral over 2D elements  $ds $ on surface $S$, giving the mutual intensity
\begin{equation}
J_{12}'=\langle \psi'_{1}(t) \psi_{2}^{'*}(t) \rangle=
\sum_{m1,n2}\langle \psi'_{m1}(t) \psi_{n2}^{'*}(t) \rangle=
\int_S ds I(s) e^{2\pi i (L_1-L_2)/\lambda}/L_1L_2
\end{equation}
where $I(s)=\langle A(t) A^{*}(t) \rangle_s$ is used to denote the ``source intensity'' at  a   patch $s$ in $S$, interpreted as a probability distribution for an initial position---  a prepared position state on the surface $S$. Here $L_1$ and $L_2$ denote distances from the surface element $ds$ to $P_1'$ and $P_2'$.
The normalized quantum-mechanical correlator in this situation is the optical ``complex degree of coherence'', which  differs from $J_{12}$ by normalizing out the   mean intensity:  $\mu_{12}'= J_{12}' / \sqrt{I(P'_1)I(P'_2)}$.
This is the appropriately normalized quantity for tracking spacetime correlations since  physical results do not depend on the ``intensity'' of the nominal carrier radiation field.

Now express the distances $L_1,L_2$ in terms of 2D coordinates in the geometrical wavefront surface.  Let
 $x,y$ denote positions of source patches in the surface $S$ and $(x_1',y_1')$ and $(x_2',y_2')$  denote positions at the points $P_1',P_2'$ in the surface $S'$, and let 
 $\Delta x'=x_1'-x_2'$ and  $\Delta y'=y_1'-y_2'$ denote position differences on $S'$. Points in each surface lie in the neighborhood of point  $O$ on $S$ and its classical null projection $O'$ on $S'$.  To leading order in small transverse displacements,
 \begin{equation}
L_1\simeq L+ { {(x_1'-x)^2+(y_1'-y)^2}\over {2L}},
\end{equation}
with an exactly similar expression for $L_2$. Thus
\begin{equation}
L_1-L_2\simeq {{(x_1'^2+y_1'^2)-(x_2'^2+y_2'^2)}\over{2L}}
-{{ x\Delta x' + y\Delta y'}\over L}.
\end{equation}
Define  
\begin{equation}
\phi  ={ {\pi [   (x_1'^2+y_1'^2)-(x_2'^2+y_2'^2)  ]}\over {\lambda L}},
\end{equation}
  corresponding to the phase difference $2\pi(OP_1-OP_2)/\lambda$ between $P_1$ and $P_2$,  relative to the classical reference ray origin $O$. (This factor can be neglected for $P_1$ and $P_2$ sufficiently close to the reference ray $OO'$,   $  (OP_1-OP_2)<<\lambda$. This applies if the width of support of $I$ is much larger than that of $\mu'$, although it does  not apply in all situations.)
Then we   find that the correlator on $S'$ is, up to a normalization,
\begin{equation}\label{transform}
\mu'(\Delta x', \Delta y') =e^{i\phi}\int_S dx dy I(x,y) \exp[{{-2\pi i(x\Delta x'+y\Delta y')}\over {\lambda L} }]
\end{equation}
This result is mathematically equivalent to the mutual coherence theory of  van Cittert and  Zernike, as described for example in \cite{bornwolf}.   
 In the quantum geometry interpretation, a  spacetime state prepared with a distribution $I$ on $S$ produces a spacetime state with a wavefunction $\mu'$ on $S'$.
The positions of events in each transverse direction  on  two  geometrical wavefronts are (apart from the phase factor $e^{i\phi}$) Fourier transforms of each other.  
This wave optics relationship resembles   that of conjugate variables in quantum mechanics.  This can be seen most clearly if we assume that $\mu'(\Delta x')$ and $I(x)$ are gaussians with respective standard deviations $\sigma',\sigma$.  These are Fourier transforms of each other obeying Eq. (\ref{transform}) if 
\begin{equation}\label{uncertain}
\sigma'\sigma=\lambda L. 
\end{equation}

The precise normalization adopted for mapping the optical model onto quantum geometry--- the choice of the Planck length $l_P$ as the fundamental wavelength $\lambda$, which sets the absolute scale for the holographic noise predictions below--- derives from   black hole physics and gravitational thermodynamics\cite{'tHooft:1993gx,Susskind:1994vu,Jacobson:1995ab,Banks:1996vh,'tHooft:1999bw,Bousso:2002ju,Padmanabhan:2006fn,Padmanabhan:2007en}.   The hypothesis adopted here for  the minimum wavelength or time of the quantum wavefront geometry, $\lambda=l_P$, will be tested with measurements of holographic noise.

  \section{holographic noise in interferometers}
   
It appears that   operational Michelson interferometric gravitational wave detectors\cite{hough} are capable  of measuring this form of quantum uncertainty   of   geometry. 
   Let $x$ denote the axis orthogonal to the incoming laser,  and normal to the wavefronts incident on the dark port.  In the language of the above analysis, the signal depends on the positions of two beamsplitter reflections in the $x$ direction, described by distributions,  $I(x)$ for one reflection and $\mu'(\Delta x')$ for the other.  The measured intensity of the light gives the phase difference of  the position at the two  reflections,
proportional to the difference in length between two arms of the interferometer.  Holographic uncertainty in the relative  positions at the two reflections  creates  noise in the measured signal, a random variation in the   position of the  beamsplitter.  The positions are compared on wavefronts with normal separation of twice the arm length $L_0$.   Let  $\sigma$ and $\sigma'$ denote the standard deviations of the the $x$ and $x'$ position wavefunctions of the beamsplitter.  We suppose that these obey the transverse position uncertainty relation (Eq. \ref{uncertain}), $ \sigma' \sigma=2 l_PL_0$, since $2L_0$ is the separation between reflection events.   The measured difference of arm lengths   of the interferometer includes a sum of both variances, which is minimized when they are equal, so $\sigma\sigma'= \sigma^2= {\sigma'}^2$.    Sampled at times farther apart than the round trip travel time, the apparent difference in arm length determined by interferometry  is a random variable with variance\cite{Hogan:2007pk}
\begin{equation}\label{variance}
\Delta L_0^2= \sigma^2+{\sigma'}^2=2\sigma^2=4 l_PL_0. 
\end{equation}
The  variation  causes   a  random  shift in the detected phase signal that appears as measurable noise.

 For the interferometric gravitational-wave detector GEO600\cite{GEO}  the noise is quoted  in terms of equivalent gravitational wave metric strain, for a wave normally incident on the detector and polarization aligned with the arms for maximum signal. The power spectral density $h^2$ gives the variance in metric strain perturbation  per frequency interval $\Delta f$.  In GEO, the arms are $L_0=600$m long and are folded once, so the time between the orthogonal beamsplitter position-measuring events is $4L_0/c$, corresponding to a resonant frequency $f_0=c/4L_0$.  
  The sum of the  holographic variance in the two directions accumulated over the resonance time is  given by Eq.(\ref{variance}):
$\Delta L_0^2=4l_PL_0$.  
The equivalent metric strain holographic  spectrum $h_H$ is
defined by equating the holographic effect on the signal with that  produced by   a gravitational wave spectrum $h_H$.
The equivalent metric strain variance is the variance of one arm (in this case, half of $\Delta L_0^2$)  divided by the squared length of the laser path along one arm (in this case, $4L_0$): 
$  {\Delta L_0^2/ 2 (4 L_0)^2}  \simeq h_H^2(f_0) \Delta f_0.$
Setting the effective bandwidth $\Delta f_0= f_0$ we obtain at $f_0= 125$ kHz,
\begin{equation}\label{flat}
h_H(f_0)= \sqrt{t_P/2}= 1.6\times 10^{-22} \  {\rm Hz}^{-1/2}.
\end{equation}
This is a factor of $\sqrt{2} $ less than the somewhat cruder estimate in \cite{Hogan:2007pk}, but the basic result remains the same: at the resonance frequency, spacetime indeterminacy produces holographic noise in this interferometer with a power spectral density  given in these units by half  the Planck time.  No parameters have been used in deriving Eq.(\ref{flat}). The  numerical value assumes that the fundamental minimum time interval is the Planck time,  $t_P\equiv l_P/c\equiv \sqrt{\hbar G_N/c^5}=5\times 10^{-44}\ \ {\rm Hz}^{-1}$.  Measurements of holographic noise will provide a direct, precision test of this hypothesis.

The  spectrum of equivalent $h_H$ in the GEO600 signal  below $f_0$ is  approximately independent of frequency,  down to the inverse residence time of a photon (about $f\simeq 550$ Hz).  The flat spectrum arises because the entire laser light power cycles many times through the two arms and the beamsplitter.  (GEO600 is more sensitive to holographic noise than LIGO\cite{Abbott:2003vs}, whose separate resonant arm cavities amplify the gravitational wave effect on the signal but not the holographic noise.)  Light reflected  from one side of the  beamsplitter at one time reaches the detected signal not just once but $\simeq f_0/f$  times.   At frequency $f$, the signal phase measures an average arm difference over a time $\simeq 1/f$, and the multiple measurements reduce the error in the mean.     As a result the holographic arm length difference  variance goes  like $\Delta L^2\propto f$,  and  $h_H$ is constant with frequency, with the value in Eq. (\ref{flat}).

The  inverse residence time for light of about 550 Hz
determines the sampling interval beyond which  the measured arm differences become  independent random variables.  At lower frequency the total displacement in phase accumulates  like a random walk, so $\Delta L^2\propto t\propto f^{-1}$,  and  $h_H\propto \Delta L/f^{1/2}\propto f^{-1}$, or
\begin{equation}\label{lowf}
h_H\simeq \sqrt{t_P/2} (f/ 550 {\rm Hz})^{-1} = 1.6\times 10^{-22} (f/ 550 {\rm Hz})^{-1} \  {\rm Hz}^{-1/2},\qquad  f< 550 {\rm Hz}.
\end{equation}
The only parameter in  the predicted holographic noise spectrum  in Eq. (\ref{lowf}) is the value of the residence time, which has been set from the measured peak frequency of the optical gain in GEO600.  The spatial and frequency dependence of the noise, as well as its absolute level, are completely fixed by the holographic geometry hypothesis.

The  predicted spectrum of holographic noise (Eqs.  \ref{flat},\ref{lowf}) approximately agrees with the spectrum of  otherwise unexplained ``mystery noise'' currently measured\cite{hild}   in GEO600 in its most sensitive band, $\simeq 300-1400$Hz, so it may already have been detected. 
 If so, the sensitivity of GEO600 is already  limited by spacetime indeterminacy. Measurements of the noise spectrum will in that case  provide a direct  and precise measurement of the fundamental maximum frequency or minimum time interval,  independent of the conventional gravitational definition of the Planck time.

 \acknowledgements
 The author is grateful for advice and information from the members of the GEO600 team,  especially S. Hild and   H. L\"uck.  This work was supported by NASA grant  No. NNX08AH33G at the University of Washington and by the Alexander von Humboldt Foundation.

\end{document}